\documentclass[prl,twocolumn,superscriptaddress,showpacs]{revtex4}
\usepackage[dvips]{graphicx}
\usepackage{latexsym}
\usepackage{bm}
\newcommand{\fig}[2]{\includegraphics[width=#1]{#2}}

\begin{document}

\renewcommand{\ni}{{\noindent}}
\newcommand{\dprime}{{\prime\prime}}
\newcommand{\be}{\begin{equation}}
\newcommand{\ee}{\end{equation}}
\newcommand{\bea}{\begin{eqnarray}} 
\newcommand{\eea}{\end{eqnarray}}
\newcommand{\nn}{\nonumber} 
\newcommand{\bk}{{\bf k}}
\newcommand{\bQ}{{\bf Q}}
\newcommand{\bN}{{\bf \nabla}}
\newcommand{\bA}{{\bf A}}
\newcommand{\bE}{{\bf E}}
\newcommand{\bj}{{\bf j}}
\newcommand{\bJ}{{\bf J}}
\newcommand{\bs}{{\bf v}_s}
\newcommand{\bn}{{\bf v}_n}
\newcommand{\bv}{{\bf v}} 
\newcommand{\la}{\langle}
\newcommand{\ra}{\rangle} 
\newcommand{\dg}{\dagger}
\newcommand{\br}{{\bf{r}}} 
\newcommand{\brp}{{\bf{r}^\prime}} 
\newcommand{\bq}{{\bf{q}}}
\newcommand{\hx}{\hat{\bf x}} 
\newcommand{\hy}{\hat{\bf y}}
\newcommand{\bS}{{\bf S}} 
\newcommand{\cU}{{\cal U}}
\newcommand{\cD}{{\cal D}} 
\newcommand{\bR}{{\bf R}}
\newcommand{\pll}{\parallel} 
\newcommand{\sumr}{\sum_{\vr}} 
\newcommand{\cP}{{\cal P}} 
\newcommand{\cQ}{{\cal Q}} 
\newcommand{\cS}{{\cal S}}
\newcommand{\upa}{\uparrow} 
\newcommand{\dna}{\downarrow}

\title{Fractionalization in Spin Liquid Mott Insulators:
Vison Wavefunctions and Gaps}
\author{Arun Paramekanti}
\affiliation{Department of Physics and
Kavli Institute for Theoretical Physics, University 
of California, Santa Barbara, CA 93106--4030}
\author{Mohit Randeria}
\affiliation{Department of Physics, University of Illinois at 
Urbana-Champaign, IL 61801}
\affiliation{Department of Theoretical Physics, Tata Institute of 
Fundamental Research, Mumbai 400 005, India}
\author{Nandini Trivedi}
\affiliation{Department of Physics, University of Illinois at 
Urbana-Champaign, IL 61801}
\affiliation{Department of Theoretical Physics, Tata Institute of 
Fundamental Research, Mumbai 400 005, India}
\begin{abstract}
\vspace{0.1cm}
We determine conditions under which a spin-liquid Mott insulator
$\vert 0\ra$ defined by a Gutzwiller projected BCS state at half-filling
is $Z_2$ fractionalized. We construct a trial vison
($Z_2$ vortex) state $\vert V\ra$ by projecting an $hc/2e$ vortex threading 
the hole of a cylinder/torus and examine 
its overlap with $\vert 0\ra$ using analytical and numerical calculations.
We find that generically the overlap vanishes in the thermodynamic limit
so the spin-liquid is $Z_2$ fractionalized.
However, for microscopic parameters appropriate for high Tc cuprates, 
we estimate the vison gap $E_v \ll J$, consistent with recent experimental
bounds, due to the proximity to the bipartite symmetric point where
$E_v = 0$.
\typeout{polish abstract}
\end{abstract}
\pacs{74.20.De,71.10.Ay,74.72.-h}

\maketitle

Spin liquid insulators doped with holes have long been regarded as
natural candidates for high temperature superconductivity since
the resonating valence bond (RVB) proposal of Anderson \cite{anderson}. 
There has been a revival of interest in this idea since 
such
wavefunctions have recently been shown to provide a 
remarkably 
good description \cite{paramekanti} of the superconducting (SC) state of the 
high Tc cuprates over a wide doping range.
In addition, spin liquid states for frustrated magnets are of
great interest both theoretically \cite{spin-liquids} and
experimentally \cite{RVBexpts}. 

It has also long been suspected that RVB insulators may exhibit
quantum number fractionalization or spin-charge separation 
\cite{kivelson,read,sachdev,wen}, i.e., 
support neutral, spin-$1/2$ and charged, spin-$0$ excitations.
Senthil and Fisher \cite {senthil_1}
have recently emphasized that such fractionalized states in $d\!\!
=\!\!2,3$ 
may be described as deconfined phases of $Z_2$ gauge theories, 
and necessarily imply a novel gapped excitation 
\cite{read,wen,senthil_1}: the $Z_2$ vortex or ``vison''.
The presence/absence of visons threading noncontractible loops
leads to a topological ground state degeneracy on cylinders or torii 
\cite{read,wen,senthil_1}. They further proposed \cite{senthil_2}
an ingenious flux-trapping experiment to detect the vison. 
However, a series of experiments to detect visons 
in highly underdoped cuprates have led to negative results, 
setting bounds on the vison gap $E_v\!\! \lesssim 150\!\!-\!\!180$K 
\cite{moler}, an order-of-magnitude smaller than the natural scale 
$J \sim 10^3$K.

Motivated by the above developments we address the following issues:
(a) How can one think about the vison in terms of {\it electronic}
coordinates?
(b) Under what conditions is the vison well-defined,
leading to $Z_2$-fractionalization and topological degeneracy 
in spin liquid insulators?
(c) How large is the vison gap, the energy scale below which fractionalization 
is apparent? (d) What are the implications for cuprates
and for other frustrated magnetic materials? 

\noindent{ \bf Vison Wavefunction:}
For conceptual clarity, consider first a square lattice of
$L^2$ sites ``wrapped'' into a cylinder along the  $\hat{x}$ direction.
The spin-liquid ground state $\vert 0 \ra$ of interest to us is
given by the Gutzwiller projection
$\cP=\prod_{\br} (1-n_{\br\upa} n_{\br\dna})$ 
of the $N$-particle d-wave BCS wavefunction at half-filling ($N = L^2$):
\be
\vert 0 \ra \equiv \cP \vert BCS \ra = \cP 
\big[ \sum_{\br,\brp} \varphi^{\vphantom\dagger}(\br - \brp)
c^{\dagger}_{\br\upa} c^{\dagger}_{\brp\dna} \big]^{N/2}
\vert {\rm vac} \ra.
\label{gs}
\ee
Here $\varphi(\rho)= L^{-2} \sum_\bk \exp(i\bk\cdot\rho) 
[\Delta_\bk/(\xi_\bk+\sqrt{\xi_\bk^2 + \Delta_\bk^2})]$
is the internal pair wavefunction with the
variational parameters \cite{notation} $\mu$ and $\Delta$
which determine $\xi_\bk = \epsilon(\bk) - \mu$,
with $\epsilon(\bk) = -2 t\left(\cos k_x + \cos k_y\right) - 
4 t^\prime \cos k_x \cos k_y$ 
and $\Delta_\bk = \Delta \left(\cos k_x - \cos k_y\right)/2$.
The $\bk$'s are chosen consistent with periodic boundary condition (PBC)
along $\hat{x}$.

With optimized variational parameters, $\cP \vert BCS\ra$ is an 
energetically good variational ground state for the 
large-$U$ Hubbard model \cite{gros,paramekanti}. However, we
focus more generally
on understanding topological degeneracy and fractionalization
in the phases
described by this {\it class} of states rather than just the energetically
optimal wavefunction. Toward this end, we define
a trial vison state $\vert V_y \ra$
by Gutzwiller projecting an $hc/2e$ vortex threading the cylinder
along $\hat{y}$ by $\vert  V_y \ra \equiv \cP \vert \left(hc/2e\right)_y \ra$,
following Senthil and Ivanov \cite{ivanov} and building on
earlier ideas \cite{footnote1}. 

To construct the $hc/2e$ vortex we simply modify the pair 
$\varphi^{\vphantom\dagger}(\br - \brp)$ in the BCS ground state
to $\exp[i\bQ\cdot(\br + \brp)/2]\varphi_{A}(\br - \brp)$, where 
$\bQ = 2\pi\hat{x}/L$ twists the phase of the condensate by $2\pi$
going around the cylinder.
However, the requirement of a {\em single-valued} wavefunction, 
when $\br$ (or $\brp$) goes around the cylinder, constrains
$\bk$'s to satisfy {\em anti}periodic boundary 
conditions (APBC) along $\hat{x}$ in the Fourier transform
defining $\varphi_{A}(\br - \brp)$; hence the subscript $A$.
For the fully projected state at half-filling one can show that
the phase factors drop out; in first quantized form, these are
independent of the spin configuration. Then the trial vison state 
$\vert V_y \ra$ simplifies to a form identical to eq.~(\ref{gs}) 
with $\varphi^{\vphantom\dagger} \rightarrow \varphi_{A}$.

Note that a projected $hc/e$ vortex with $\bQ = 4\pi\hat{x}/L$ 
does not require any change in boundary conditions on $\varphi$
and is therefore identical to the ground state $\vert 0 \ra$.
The manifest $Z_2$ character (two visons is the same as none) 
of projected vortices makes them attractive candidates for 
visons. On a torus $\vert 0 \ra$ corresponds to PBC 
along both $\hat{x}$ and $\hat{y}$, the vison states $\vert V_x \ra$ and 
$\vert V_y \ra$ correspond to PBC/APBC
and $\vert V_{xy} \ra$ to APBC along both directions.

\medskip

\ni{\bf Phase diagram and Symmetries}:
A superconducting state with a vortex threading the hole of
the cylinder/torus carries
current and is orthogonal to the ground state.
However, Gutzwiller projecting the BCS wavefunction at 
half-filling destroys SC order \cite{paramekanti} and results in a 
spin-liquid insulator \cite{footnote2}. The question we would like
to address is: {\it does any remnance of vorticity survive projection?}
If it does, then
$\la V_\alpha \vert 0 \ra\!\! =\!\! \la V_\alpha \vert V_\beta \ra
\!\!= \!\!0$,
for $\alpha,\!\!\beta\!\! =\!\! x,y,xy$, the vison is well-defined
leading to topological degeneracy
\cite{footnote3}, 
and one is in a $Z_2$ fractionalized
phase. If, on the other hand, the proposed vison state has nonzero 
overlap with the state $\vert 0\ra$ in the thermodynamic limit,
one is in a conventional non-fractionalized phase.
Our main task then is to compute these overlaps
and establish the existence of disconnected topological sectors 
as evidence for fractionalization.

\begin{figure}
\begin{center}
\vskip-2mm
\hspace*{0mm}
\centerline{\fig{3.0in}{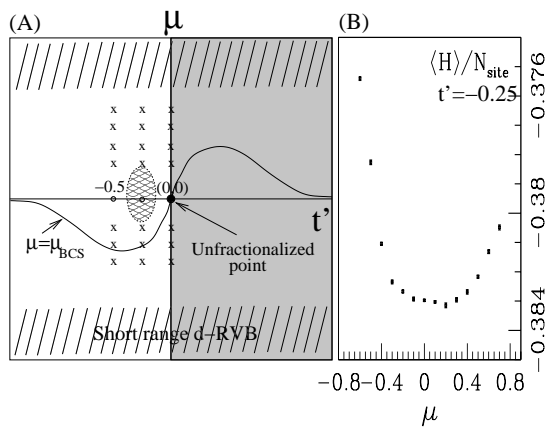}}
\vskip-6mm
\caption{(A) ``Phase diagram'' showing that $\cP\vert 0\ra$ is $Z_2$ 
fractionalized in the entire $(t^\prime,\mu)$-plane except
at the bipartite symmetric point $(0,0)$ (see text for details). (1) 
Shaded and unshaded half-planes
are related by symmetry $(t^\prime,\mu) \to (-t^\prime,-\mu)$;
(2) hatched region for large negative $\mu$ is the 
short-range RVB limit; (3) analytical arguments for
$\mu=\mu_{\rm BCS}$ (schematic line) are described in the text;
(4) crosses indicate some points where we
numerically compute overlaps; (5) shaded ellipse around $(-0.25,0)$ 
is the region relevant for high Tc SCs. 
(B) Energy of $\vert 0 \ra$ for the
Hubbard model at half-filling for parameters relevant to cuprates 
($U/t\!\! =\!\! 12, t^\prime/t\!\! =\!\! -0.25$ and optimal $\Delta\!\! 
=\!\! 1.25t$) versus $\mu/t$. The optimal value is
$- 0.3 \lesssim \mu/t \lesssim 0.3$ within statistical errors.}
\label{fig1}
\end{center}
\vskip-8mm
\end{figure}

To determine the conditions under which the vison survives, 
it is useful to consider the full parameter space for 
projected BCS states at half-filling, which is the  
$(t^\prime,\mu)$ plane with $t = 1$ and 
$\Delta$ held fixed at some non-zero value
in order to describe a RVB liquid of singlet pairs.

We can use symmetry arguments to show that not all
states in this space are distinct upon projection and
$\vert 0 (t^\prime,\mu)\ra = \vert 0 (-t^\prime,-\mu)\ra$. 
In brief, we can change $t^\prime\to -t^\prime$, $\mu \to 
-\mu$ by a global particle-hole transformation 
$c^{\vphantom\dagger}_{\br\sigma} \to (-1)^{x+y} c^{\dagger}_{\br\sigma}$ 
in the wavefunction, also redefining the $\vert {\rm vac} \ra$ since 
empty sites transform into doubly occupied ones. However, with 
{\it exactly} one particle per site such a transformation
interchanges $\upa$-spins and $\dna$-spins at every site for any 
configuration. Since the BCS wavefunctions are spin-singlets this 
leaves the state invariant. We thus restrict attention to 
$t^\prime \le 0$ in Fig.~\ref{fig1}(A).

\medskip

\ni{\bf Short-range RVB limit:}
Note that large negative $\mu$ with 
$\mu < - 4 (\vert t\vert +\vert t^\prime\vert)$
is the bosonic limit \cite{randeria} of the BCS wavefunction, with
$\varphi(\br -\brp)$ decaying exponentially in real space. 
In this limit $\vert 0 \ra$ may be viewed as a short-range RVB state.
Earlier studies of closely related dimer models \cite{kivelson,
read} indicate that, for short-range singlet bonds there exist 
four topological sectors on a torus, which may 
be straightforwardly related to sectors with/without
visons in the $\hat{x}$, $\hat{y}$ directions.
Further, two dimer {\it configurations} from 
different sectors have exponentially small overlap $\sim \exp(-\alpha L)$.
A state is a superposition of many configurations, and {\em assuming} that 
individually small overlaps do not add up coherently,
the overlap of two states from different sectors would vanish
as $L \to \infty$. 

We have numerically checked (see below) that the overlap
vanishes for large $L$ validating the above assumption, 
and implying fractionalization in the short range RVB limit.
Using the symmetry discussed above, we conclude that
states in the region with (positive)
$\mu > 4(\vert t\vert +\vert t^\prime \vert)$
are also fractionalized, even though $\varphi(\br- \brp)$ is 
{\it not} obviously short-ranged here.

\medskip

\ni{\bf Analytical insights for $\mu=\mu_{\rm BCS}$:} 
We next focus on states for which $\mu=\mu_{\rm BCS}(t^\prime)$ 
(see the curve in Fig.~\ref{fig1}(A)),
where $\mu_{\rm BCS}$ is the value at which the unprojected BCS
state is at half-filling. The {\em un}projected state may then be 
viewed as the ground state of a d-wave BCS mean field (MF) Hamiltonian and
it is a coherent superposition of number eigenstates, sharply peaked 
at the correct mean density. Gutzwiller projection picks out 
the $N = L^2$ contribution with no double occupancy. 

Define local $SU(2)$ gauge 
transformations \cite{affleck} ${\cU}$, generated by
$T_i^+= c^{\dagger}_{i\upa} c^{\dagger}_{i\dna}$,
$T_i^-= c_{i\dna} c_{i\upa}$,
$T^z_i=\sum_{\sigma}c^{\dagger}_{i\sigma}c^{\vphantom\dagger}_{i\sigma} -1$, 
which mix empty and doubly occupied
sites as an $SU(2)$ doublet, but act trivially
on the projected subspace with {\it exactly} one particle per site.
Gutzwiller projection is then equivalent to projection onto
the $SU(2)$ singlet subspace, and we may write any state
$\cP \vert \Phi \ra = \int_\cU \cU \vert \Phi \ra$,
where the integral is over all group elements ${\cU_\theta}=
\exp(i\sum_i \vec{T}_i\cdot\vec{\theta}_i)$.
The overlap of interest can then be written as
\be
\la V \vert 0 \ra = \int_\cU \la (hc/2e) \vert \cU \vert BCS \ra,
\label{overlap}
\ee
which now reduces the problem to computing overlaps of
{\em un}projected states.

For a nonbipartite lattice (e.g., with $t^\prime \ne 0$),
one cannot gauge away the off-diagonal term in the
d-wave BCS MF Hamiltonian and $\cU\vert BCS \ra$ 
is a SC state (or vortex vacuum) for arbitrary $\cU$ \cite{ivanov}. 
Then $\la (hc/2e) \vert \cU \vert BCS \ra =0$ for arbitrary 
$\cU$ and $\la V \vert 0 \ra$ vanishes.
Thus non-bipartiteness is a sufficient condition for
fractionalization in d-wave states when $\mu=\mu_{\rm BCS}$ 
\cite{previous}.

On the other hand, for a bipartite system
(in our case $t^\prime = 0$), it is well known \cite{affleck}
that one can gauge away the off-diagonal part of the 
MF Hamiltonian (provided it doesn't break time reversal) and
transform the BCS state into a staggered flux state.
Now $\la (hc/2e) \vert \cU \vert BCS \ra$ is non-zero
for some choice $\cU$ and one cannot use eq.~(\ref{overlap})
to argue that $\la V \vert 0 \ra$ vanishes.

\medskip

\ni{\bf Overlaps within $Z_2$ gauge theories:} 
Before proceeding to the numerical results, it is useful to 
understand how overlaps like $\la V \vert 0 \ra$ are expected
to scale with system size $L^2$. In a non-fractionalized phase
the overlap will remain non-zero as $L \to \infty$.
However, in a fractionalized phase we argue that it should 
vanish exponentially for large $L$.
This is easy to see when the matter fields are gapped 
(e.g., the short-range RVB limit) and may be
integrated out to obtain a $2+1$ dimensional $Z_2$ gauge theory.
Deep in the deconfining phase of a pure gauge theory,
we have evaluated the overlap perturbatively and it scales as 
$\la V \vert 0 \ra \sim \exp(-L/\xi)$. To understand this
result note that the overlap in the deconfined phase behaves 
similar to the spin-spin correlation
of the dual Ising model at distance $L$, where the dual
spins represent the gapped visons. Since the dual Ising model 
is disordered when the gauge theory 
is deconfining, the exponential decay follows. 
Another way to understand 
this result is that the overlap is nonzero due to
tunneling of the vison (a ``string'' of length $L$) out of the
hole of the cylinder. Such tunneling events involve intermediate
states in which the vison pierces the surface of the cylinder.
This costs an energy proportional to the vison gap, and leads
to an overlap exponentially small in $L$.
The situation with gapless matter fields is less well studied,
even though numerically (see below) we find clear evidence
for exponential decay. 

\medskip

\ni{\bf Numerical results:}
We finally discuss the results of numerical evaluation of vison overlaps 
using the variational Monte Carlo method.
For computational simplicity \cite{numerics}
we focus on the overlap $\la V_x|V_y\ra$.
We choose $t=1$ and fix $\Delta = 1.25$ for all the
calculations reported below.
The following numerical results are obtained in regimes where
we have given analytical arguments above.
(1) We have explicitly checked that the 
overlaps are invariant under
$(t^\prime,\mu) \to (-t^\prime,-\mu)$. 
(2) In the regime of short-range RVB
we find that the system is $Z_2$ fractionalized; see Fig.~2(A)
for large negative $\mu$. 
(3) On the curve $\mu=\mu_{\rm BCS}$ we find
that the non-bipartite systems are fractionalized. (4) For
the special point $(t^\prime,\mu) =(0,0)$ with
bipartite symmetry, we indeed find that the overlaps are non-zero,
independent of $L$ (see Fig.~2(A)), the 
vison is not well defined and the system is
{\em not} fractionalized. 
Although consistent with
the analytic arguments, the numerical results are important because
the arguments were not rigorous for (2) and (3), and
silent on point (4).

\begin{figure}
\begin{center}
\vskip-2mm
\hspace*{0mm}
\centerline{\fig{3.5in}{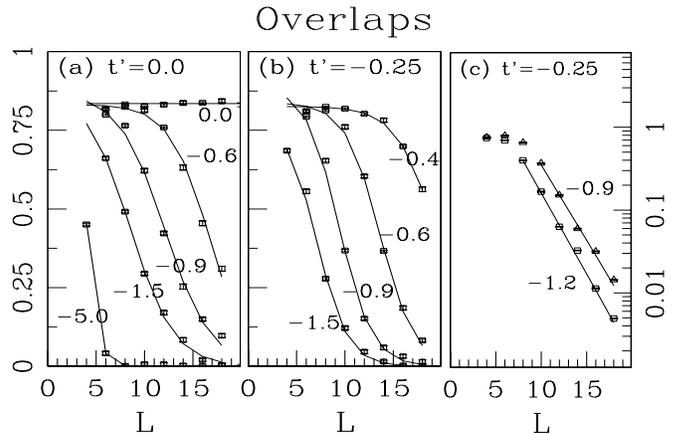}}
\vskip-6mm
\caption{Size dependence of the overlap $\la V_x|V_y\ra (L)$ 
for $\Delta = 1.25$ with
(A) $t^\prime = 0$ and (B) $t^\prime=-1/4$. The various curves
correspond to the $\mu$ value indicated.
The lines are fits
to the form $\la V_x|V_y\ra(L)=a (1-\tanh((L-\xi^\ast)/\xi))$,
which works well for the entire regime of parameters we have studied,
and from which we extract $\xi^\ast$. (C) Exponential asymptotic
behavior of the overlap shown on semi-log scale.
}
\label{fig2}
\vskip-8mm
\end{center}
\end{figure}

We next evaluate overlaps at $t^\prime =0.0,-0.25,-0.5$ 
for $-1.5 \lesssim \mu \lesssim 1.5$ which, as we
discuss below, covers the region of interest for possible spin-liquid
insulators in the vicinity of the high Tc cuprates.
We show in Fig.~\ref{fig2} the overlap $\la V_x \vert V_y \ra$
as a function of system size $L$ for a range of $t^\prime$ and 
$\mu$. We find that at small $L$ the 
overlaps are finite and then cross over on a scale $\xi^\ast$ 
to an asymptotic decay $\exp(- 2 L/\xi)$. This behavior can
be simply described by the functional form 
$\la V_x|V_y\ra (L)=a(1- \tanh((L-\xi^\ast)/\xi))$
which appears to fit the data well.
(The factor of two is included in the asymptotic decay
definition since we are looking at overlap of states with two
visons in different directions). In the regime of parameters where 
we can reliably extract both $\xi^\ast$ and $\xi$, we find
$\xi^\ast \simeq 2 \xi$. More generally, it is hard to access the
asymptotic behavior in the region of interest, and we can extract
crossover scale $\xi^\ast$ more accurately than $\xi$. 

\begin{figure}
\begin{center}
\vskip-2mm
\hspace*{0mm}
\centerline{\fig{2.5in}{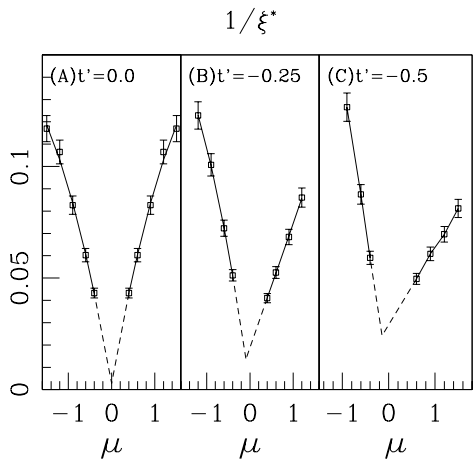}}
\vskip-6mm
\caption{Dependence of $1/\xi^\ast$ on
$\mu$ for (A) $t^\prime=0.0$, (B) $t^\prime=-0.25$, 
(C) $t^\prime=-0.5$. In panel (A), $1/\xi^\ast$ extrapolates to zero 
on approaching $(t^\prime,\mu)=(0,0)$ indicating a
non-fractionalized ``singular point'' at the origin. 
For $t^\prime=-0.25,-0.5$ (B,C) the length scale $\xi^\ast$
appears finite everywhere.
}
\label{fig3}
\vskip-8mm
\end{center}
\end{figure}

We plot the inverse length scale $1/\xi^\ast$ for $t^\prime=0,-0.25,-0.5$ at 
different $\mu$ in Fig.~\ref{fig3}. For $t^\prime=0$, we see that
$\xi^\ast \to \infty$ as $\mu \to 0$,
fully consistent with the finite overlap independent of $L$ on 
accessible system sizes for $(t^\prime,\mu) = (0,0)$ seen 
in Fig.~\ref{fig2}(A). It also clear that $\xi^\ast$ remains finite 
everywhere away from the origin though it may become quite large in 
its vicinity.

Since fractionalization manifests itself at length scales 
larger than $\xi^\ast$, our numerical results strongly suggest 
that all points in the phase diagram of Fig.~\ref{fig1}(A) are 
fractionalized, except for the origin where $\xi^\ast$ diverges. 
$(t^\prime,\mu) = (0,0)$,
with its special bipartite symmetry, is a non-fractionalized 
``singular point'' in the space of spin-liquid insulators that we 
study.

\medskip

\ni{\bf Vison gaps:}
Now the question arises: what regime in parameter space
is relevant for the cuprates? One possibility is that there
are no fractionalized states in the vicinity of
the observed phases in real materials. An alternative
worth exploring in view of the success of
RVB wavefunctions in understanding the SC state \cite{paramekanti},
is that we take the insulating limit (hole doping $x\!\!\to\!\!0$) of
$\cP\vert BCS\ra$. Optimizing variational parameters 
$\mu,\Delta$ for the large-$U$ Hubbard model ($U\!\!=\!\!12, 
t^\prime\!\!=\!\!-1/4$),
we find $\Delta\!\!\simeq\!\!1.25$ while $-0.3\lesssim \mu \lesssim 0.3$ as 
shown in Fig.~\ref{fig1}(B). For this region (the shaded ellipse in 
Fig.~\ref{fig1}(A)) we conclude from Fig.~\ref{fig3}(B) that
$\xi^\ast \gtrsim 25$ lattice spacings. 

We now convert this to an estimate of the vison gap $E_v$.
We expect $E_v= \alpha J/(\xi^\ast)^z \le \alpha J/\xi^\ast$ 
which vanishes at the ``singular point'' with bipartite symmetry.
Here $J$ is the nearest-neighbor superexchange, $\alpha\equiv 
{\cal O}(1)$ is a dimensionless constant and the dynamical exponent 
$z \ge 1$.
For $\alpha\!\!=\!\!1$ and $J\!\!=\!\!1200 K$, the estimated 
$E_v \lesssim 50$K (but this energy
could be higher if $\alpha$ is larger).
More concretely, we emphasize that 
proximity to the `bipartite point' $(t'=0,\mu=0)$ can lead to a very small 
vison gap for fractionalized RVB states. 
This is consistent 
with existing experimental bounds \cite{moler} and suggests 
that fractionalization is apparent only at 
very low $T$ and does not play a role in pseudogap anomalies. 

The ideas and methods developed here should be explored further
in highly frustrated magnets,  
with large $t^\prime$ or ring-exchange 
interactions \cite{ring-exchange}, which may be more promising candidates 
for observing fractionalization.

\medskip

\ni{\bf Acknowledgments:} 
We thank T. Senthil for extensive discussions and for
generously sharing his ideas and unpublished notes with us.
We thank also L. Balents, D.~M. Ceperley, M.~P.~A. Fisher, 
E. Fradkin and A. Melikidze 
for helpful discussions. MR and NT gratefully acknowledge the 
hospitality of the Physics Department at University of Illiois and
support through DOE grant DEFG02-91ER45439 and DARPA grant N0014-01-1-1062.
AP was supported by NSF DMR-9985255 and PHY99-07949 and
the Sloan and Packard foundations. We acknowledge the use of computational 
facilities at TIFR including those provided by the DST Swarnajayanti 
Fellowship.


\begin{thebibliography}{999}
\bibitem{footnote2} 
We have checked that these wavefunctions have neither long-range spin order 
nor dimer order and thus describe translationally invariant spin-liquids.
\bibitem{footnote3}
We have checked that local spin 
correlations are identical in the thermodynamic limit for states with 
and without a vison. These states are thus locally identical 
and differ only in their global (topological) properties.
\bibitem{numerics} The Fourier transform of the d-wave
$\varphi(\bk)$ has singularities on the zone diagonal.
Thus on the square lattice we can only define 
$\vert V_x \ra$ and $\vert V_y \ra$. We have also studied
other geometries which permit
all four states $(0,V_x,V_y,V_{xy})$; the results
will be described elsewhere.
\bibitem{previous}
The same result was obtained by Wen \cite{wen} by analyzing stability of 
mean field solutions for spin models.
\end{thebibliography}
\end{document}